\documentclass[12pt,letterpaper]{article}
\usepackage[margin=1in]{geometry}
\usepackage{color}
\usepackage{comment}
\usepackage[utf8]{inputenc}
\usepackage[numbers]{natbib}
\usepackage{graphicx}
\usepackage{amsmath}
\usepackage{amssymb}
\usepackage{hyperref}
\usepackage{lineno}
\setcitestyle{square}

\def\apjl{ApJL}                
\def\apjs{ApJS}               
\def\aap{A\&A} 

\newcommand{\snia}{Type Ia SN}

\newcommand{\h}{\ensuremath{\mathrm{H}_0}}
\newcommand{\lcdm}{\ensuremath{\Lambda}CDM}
\newcommand{\mpckms}{\ensuremath{\mathrm{km}/\mathrm{s}/\mathrm{Mpc}}}

\usepackage{sectsty}
\sectionfont{\large}
\subsectionfont{\normalsize}
\subsubsectionfont{\normalsize}
\paragraphfont{\normalsize}

\begin{document}

\pagenumbering{gobble}
\huge
\begin{center}
Astro2020 Science White Paper
\vspace{0.1in}

Single-object Imaging and Spectroscopy to Enhance Dark Energy Science from LSST
\end{center}
\normalsize

\noindent\textbf{Thematic Areas:} Cosmology and Fundamental Physics\\

\noindent\textbf{Principal (corresponding) author:}\\
Name: R.~A.~Hlo\v{z}ek \\
Institution: Department of Astronomy and Astrophysics \&  Dunlap Institute for Astronomy and Astrophysics, University of Toronto\\
Email: hlozek@dunlap.utoronto.ca\\
Phone: +1 416-978-4971 \\

\noindent\textbf{Co-authors:} T.~Collett (IoCG, Portsmouth), L.~Galbany (U. Pittsburgh, PITT PACC), D.~A.~Goldstein (Caltech, Hubble Fellow), S.~W.~Jha (Rutgers), A.~G.~Kim (LBNL), R.~Mandelbaum (CMU), J.~A.~Newman (U. Pittsburgh, PITT PACC), S.~Perlmutter (LBNL, UC Berkeley), D.~J.~Perrefort (U. Pittsburgh, PITT PACC), M.~Sullivan (Southampton), and A.~Verma (U. Oxford),  for the LSST Dark Energy Science Collaboration\\

\noindent\textbf{Abstract:} Single-object imaging and spectroscopy on telescopes with apertures ranging from $\sim 4$m to $40$m have the potential to greatly enhance the cosmological constraints that can be obtained from LSST.  Two major cosmological probes will benefit greatly from LSST follow-up: accurate spectrophotometry for nearby and distant Type Ia supernovae will expand the cosmological distance lever arm by unlocking the constraining power of high-$z$ supernovae; and cosmology with time delays of strongly-lensed supernovae and quasars will require additional high-cadence imaging to supplement LSST, adaptive optics imaging or spectroscopy for accurate lens and source positions, and IFU or slit spectroscopy to measure detailed properties of lens systems.  We highlight the scientific impact of these two science drivers, and discuss how additional resources will benefit them. For both science cases, LSST will deliver a large sample of objects over both the wide and deep fields in the LSST survey, but additional data to characterize both individual systems and overall systematics will be key to ensuring robust cosmological inference to high redshifts.  Community access to large amounts of natural-seeing imaging on $\sim$2--4m telescopes, adaptive optics imaging and spectroscopy on 8--40m telescopes, and high-throughput single-target spectroscopy on 4--40m telescopes will be necessary for LSST time domain cosmology to reach its full potential.  
In two companion white papers we present the additional gains for LSST cosmology that will come from deep and from wide-field multi-object spectroscopy. 


\newpage
\pagenumbering{arabic}

\section{Introduction}

As a Stage IV Dark Energy program, the Large Synoptic Survey Telescope (LSST) will play a major role in improving our knowledge of cosmology over the years 2023--2033 via both wide-area and deep surveys. However, obtaining measurements that extract the full potential of LSST will require additional data from other ground-based facilities.  
In this white paper, we describe the science opportunities to build on and strengthen the LSST dataset that would be made possible by community access to telescopes and instruments enabling detailed follow-up of individual objects. These facilities can provide single-object imaging either in more bands, with higher spatial resolution, or with higher cadence than the LSST observations provide, and can provide spectroscopic measurements for rare objects, which are difficult to target efficiently for multi-object spectroscopy.  

In two companion white papers we describe the gains for LSST cosmology that would come from community access to highly-multiplexed optical and infrared spectroscopy on 4--40~m telescopes, either via surveys of faint objects at the full depth of LSST cosmological samples, i$\sim$25 \cite{deep}, or from wider-area ($>20$deg$^2$) but shallower surveys \cite{wide}. 


\section{Type Ia Supernova Spectroscopy to High Redshifts}

As described in a companion paper \cite{wide}, the largest sets of redshift measurements for LSST supernovae should come from targeting SNe and their host galaxies via wide-field multi-object spectroscopy, either using a small fraction of fibers in surveys that span the LSST footprint or via targeted surveys in the LSST Deep Drilling Fields (DDFs).  However, higher signal-to-noise observations will be needed for a significant sample of low-$z$ SNe to characterize the range of intrinsic spectra; and to compare this to the high-$z$ LSST \snia\ ($z \sim 0.5 - 1.4$) will require targeted long-exposure follow-up while the high-$z$ SNe are still bright.

\textbf{Typing to eliminate non-SNe~Ia contamination and to validate photometric classification at high (and low) redshift:}  Much work has gone into techniques for determining supernova types and redshifts from photometry alone, but this is still a difficult technique when applied over a large redshift range due to changing filter coverage and unknown drifts in the spectroscopic properties of the SNe.  Current photometric identification techniques require spectroscopically classified training samples, ideally over the full range of redshifts under consideration.  Fully characterizing some classes of objects from LSST at the faintest magnitudes \citep[e.g., SLSN and Type II SN, see][]{deJaeger2017, inserra2018} will require spectroscopic observations on 25--40~m class telescopes. Furthermore, measuring the redshifts via spectra of ``live'' supernovae eliminates type uncertainty, yielding high-purity samples with limited systematics. Ultimately, detailed spectrophotometric studies of low-$z$ \snia~will be the best source of high-signal-to-noise information about the behavior over a large range of \snia~subtypes, making spectroscopy on smaller telescopes valuable as well. An example of such a study can be found in \cite{Saunders2018}.

\textbf{Systematic error mitigation:}
Detailed understanding of a variety of potential sources of systematic error will be crucial for supernova science with LSST, including evolution of the \snia\  population with redshift; errors in the specification of the spectral energy distribution (SED) in different redshift/rest-frame wavelength ranges that can affect training of light curve models; and mis-classifying fine sub-types (``twins'') of Type Ia supernovae. 
Finally, we do not know if our current catalog of spectral properties is exhaustive, and, if not, the mismatch between assumed spectral model templates and actual spectra will introduce systematic shifts in our cosmological distance measurements. While deep follow-up of a variety of LSST transients will be needed to understand their physics and evolution, requirements for the use of SNe Ia as cosmic probes of distance are particularly stringent, and hence detailed spectral studies are needed.

Very high data quality is needed if spectra of ``live" supernovae will be used to detect the subtle differences between sub-types of SN Ia, to study evolutionary population drift, and to match highly-similar supernovae (``twins'' as described in \cite{Fakhouri2015}) to reduce the dispersion of measured distances. So far, such purposes require high signal-to-noise observations to track subtle spectroscopic features and spectrophotometric data to allow a clean subtraction of the host galaxy light.  For systematics constraints on measurements of high-redshift \snia\ to be commensurate with the precise statistical uncertainties resulting from large numbers of LSST-discovered SNe, there will be a need for new instrumentation for ground-based telescopes (e.g., IFU or high-throughput spectrographs now in the planning stages for 2--4~m telescopes and IFU reformatters on existing spectrographs (with or without AO) on 8--40~m telescopes), as well as development of space-based observational avenues such as coordinated observations with Euclid, WFIRST or other space missions now planning spectrophotometric instrumentation.


\section{Cosmology with Strong Gravitational Lens Systems}
One of the most striking consequences of general relativity is that light from distant sources is deflected by the gravitational field of massive objects near the line of sight.
When the deflection produces multiple images, the phenomenon is known as ``strong gravitational lensing.'' 
The multiple images of strongly lensed sources arrive at different times because they  travel different paths and through different gravitational potentials to reach us \citep{refsdal64a,blandfordandnarayan92}.
When a strongly lensed source is time variable, the arrival time delays can be measured and combined with a mass model to yield cosmological constraints, particularly on the Hubble constant \h.  Strongly-lensed QSOs, compound lens systems,  and strongly-lensed supernovae each provide opportunities for measurement of cosmological parameters with LSST, and the LSST DESC plans to exploit them all \citep{descsrd}.

\subsection{Precision Cosmology with Strongly-Lensed Supernova Time Delays}

\begin{figure}[htbp!]
	\centering
    \includegraphics[width=0.5\textwidth]{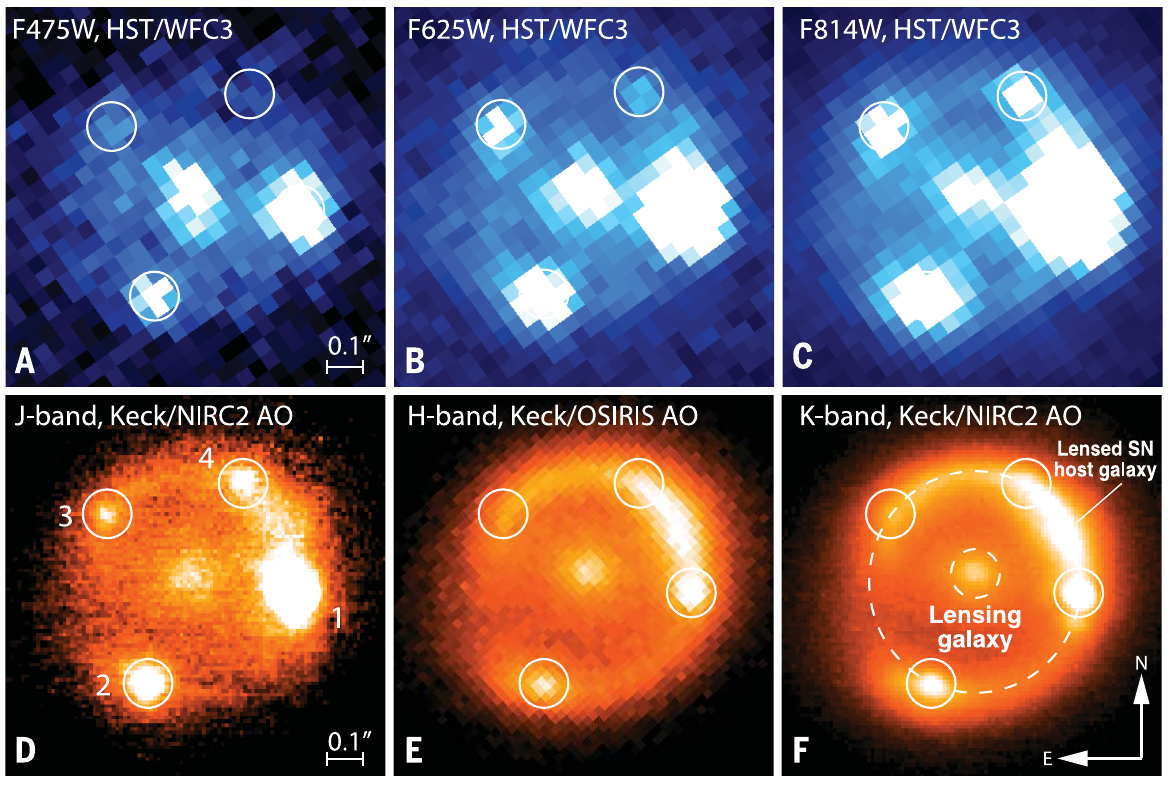}
    \caption{iPTF16geu, a Type Ia supernova at $z=0.4$ strongly lensed by an elliptical galaxy at $z=0.2$ (data from \textit{HST}\ and Keck). 
    This transient---the only known lensed \snia\ with resolved multiple images---was discovered by a predecessor of LSST, the intermediate Palomar Transient Factory (iPTF).
    The multiple images of the supernova are indicated with circles. 
    Due to the small image separation ($\sim$0.3$^{\prime\prime}$), high-resolution imaging was required to confirm that it was a strongly-lensed supernova.
    Figure reproduced from \cite{goobar16}.}
    \label{fig:geu}
\end{figure}

The prospect of cosmological constraints from strongly-lensed supernovae was  illustrated with the ground-breaking Hubble Space Telescope observations of the first multiply-imaged lensed supernova, SN Refsdal \cite{kelly}. Ground-based time-domain optical imaging surveys similar in spirit to LSST, such as the intermediate Palomar Transient Factory (iPTF), played a key role in the discovery and follow-up observations of another strongly gravitationally lensed supernova with resolved multiple images \citep[][see Figure \ref{fig:geu}]{goobar16}.

Time delays from lensed supernovae present opportunities to observe the earliest phases of supernova explosions, to infer cosmological parameters, and to map substructure in lens galaxies, but many more systems are needed to achieve these goals. Thanks to novel lensed-supernova hunting techniques, a new generation of alert-based wide-field imaging surveys that  began in mid--2018 with the Zwicky Transient Facility \citep[ZTF;][]{ztfnat} and will continue into the 2030s with LSST  will increase the size of this sample by orders of magnitude. 
ZTF is currently expected to yield $\sim$20 lensed supernovae over the course of its 3-year campaign, followed by thousands from LSST and WFIRST \citep{gnkc17}. 
This vast increase in sample size will enable groundbreaking new measurements with the potential to rapidly deliver precision constraints on the Hubble constant (\h) and dark energy.  The Hubble constant can be determined to exquisite precision with measurements of the \textit{nearby}\ universe  using the distance ladder \citep[][$\h=73.24\pm1.74\,\mpckms$]{riess16}. 
It can also be inferred with measurements of the \textit{primordial}\ universe using the cosmic microwave background (CMB), assuming a \lcdm\ cosmology \citep[][$\h=66.93\pm0.62\,\mpckms$]{planckhitens}. 
The tension between these local and distant measurements is palpable: they currently disagree by $3.8\sigma.$ 
It is potentially a sign of new fundamental physics, such as sterile neutrinos or ``phantom'' dark energy \citep[e.g.,][]{phantom,freedman17,zhao17}, but could also be a sign of systematics in the measurements \citep[e.g.,][]{efstathiou14,trouble}.
Time delays between the multiple images of strongly lensed time-variable sources depend primarily on \h, and the mass distribution along the line of sight \citep{kochanek02, treu2010,fassnacht11,cosmography}.
Refsdal \cite{refsdal64b} first suggested using time delays from strongly lensed supernovae to  measure \h, but today, more than 50 years later, this measurement has yet to be made with precision.
Time delays from lensed supernovae have many advantages over time delays from strongly lensed quasars, which have been used to measure \h\ to 3.4\% in a \lcdm\ cosmology 
Lensed supernovae require 100 times less monitoring  (i.e., a few weeks rather than decades) and are less sensitive to microlensing \citep{gnkc17},  mass modeling systematics \citep{oguri03}, and selection bias \citep{collettcunnington16}.
Consequently, they provide the most direct and rapid route to sub-percent constraints on \h\ with strong lensing time delays.

\begin{figure}[htbp!]
	\centering
    \hspace{-37pt}
    \includegraphics[width=0.6\textwidth]{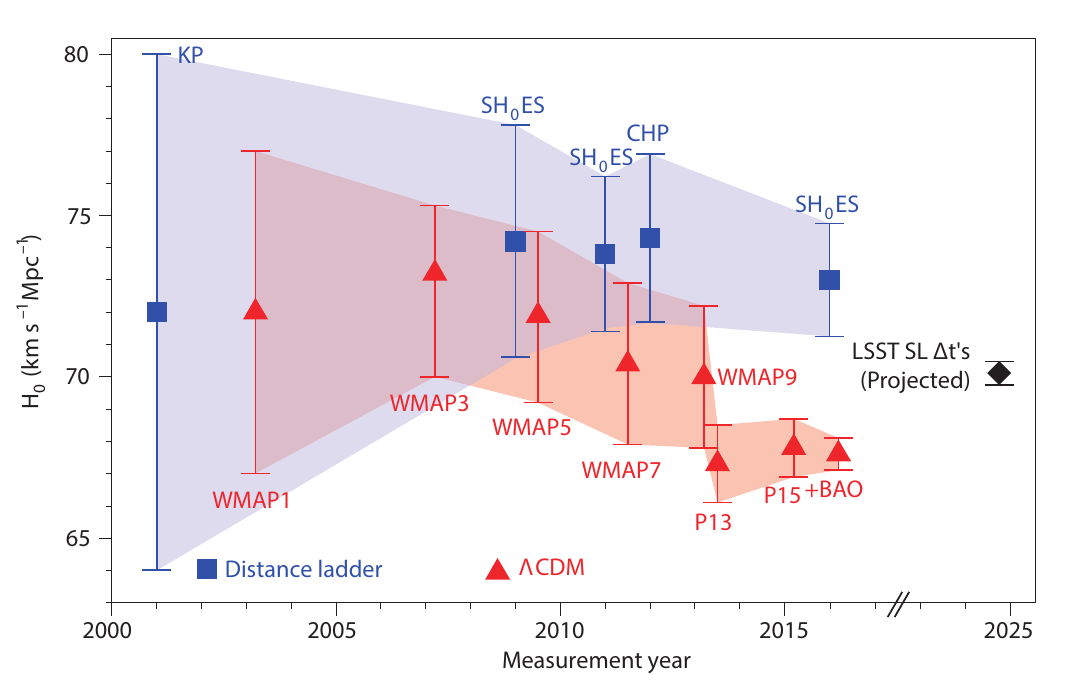}
    \caption{
    \footnotesize Examples of measurements of \h\ since the year 2000. Strong lensing time delays from systems discovered by LSST and observed with supplemental monitoring and follow-up spectroscopy as recommended in this white paper (black diamond) will resolve or increase the current tension in \h.
    Figure adapted from \cite{freedman17}.}
    \label{fig:impact}
\end{figure}

We forecast that \h~can be measured to sub-percent precision within \lcdm\ cosmological models with time delays from systems discovered in the first year of the LSST survey. The total number of supernova time delay systems over the LSST survey is expected to be $\sim 100$. Our projected constraint on \h\ as shown in Figure~\ref{fig:impact} is comparable in  precision to the leading current measurement from the combination of $\textit{Planck}$ and BOSS data \citep{aubourg15,planck15}, and it is almost ten times better than the current state-of-the-art constraints from quasar time delays \citep{bonvin17}. In addition to constraining \h, time delays from lensed supernovae are sensitive to dark energy in a completely different way than cosmological probes based on distances and volumes \citep[e.g., the CMB, the \snia\  distance-redshift relation, BAO, and galaxy clusters;][]{linder04,linder11}, making time-delay measurements highly complementary. Adding in lensed supernovae from the first year of LSST increases the Dark Energy Task Force \cite{detf_report} figure of merit by a factor of 3 over an \snia-based constraint alone---a major gain.

\subsection{Photometric and Spectroscopic Follow-up of Lens Systems}

LSST is expected to deliver hundreds of cosmologically useful lensed supernovae \citep{2018arXiv180910147G} and thousands of cosmologically-useful lensed QSOs \citep{om10}. Initial observations of strong lenses piggybacking on wide-field multi-object spectroscopic surveys can help to characterize lensed QSO systems and identify the most useful ones, as described in a companion paper \cite{wide}.
However, measuring cosmological parameters to percent-level accuracy with strong-lens time delays requires two main ingredients once the ideal systems are identified. First, one must measure the time delays between the multiple images of a source, and second, the lensing potential must be inferred to convert the observed time delays into a time delay distance. 

To measure time delays, high-cadence, high-resolution, multi-filter imaging of the resolved lensed images is required. In general, the LSST cadence may be insufficient for this purpose; as a result, dedicated follow-up imaging of lens systems with more frequent visits will be needed, but this can often be performed on smaller telescopes (2--4~m).
To model the lens systems, both  source and lens redshifts are required, as well as high-resolution imaging of the lens galaxy and lensed host galaxy to measure apparent positions with high precision. Kinematic velocity dispersions derived from lens galaxy spectra can also improve the models.  Adaptive optics (AO) IFU spectroscopy on 8--40m-class telescopes (with aperture required depending on brightness) can satisfy all of these needs simultaneously, but the combination of AO imaging with slit spectroscopy would also suffice.

\section{Recommendations}

Important advances in cosmological studies with SNe and strongly lensed systems in the new era of large-area, high-cadence optical surveys such as LSST will require single-object follow-up data of several types:

\begin{itemize}
    \item Spectrophotometry of SNe Ia will enable direct cosmological measurements, the development of spectral templates and the calibration of photometric classification performance at high redshift (where it is the most uncertain), as well as tests of systematic effects. Given the wide range of LSST SN brightnesses, this work will be reliant on large amounts of low-to-medium resolution spectroscopy on ground-based telescopes with apertures from 4 m to $>20$ m, and, ideally, space-based telescopes as well.
    \item Follow-up imaging with more frequent cadence than LSST will be valuable for measuring strong lensing time delays; this can generally be conducted on 2--4~m telescopes.
    \item Adaptive optics IFU spectroscopy, or AO imaging plus slit spectroscopy, is needed for precision image position measurements and lens system modeling; this work will require access to both 8--10~m and $>30$ m telescopes with AO capabilities.
\end{itemize}
The Kavli/NOAO/LSST report (summarized in Table~\ref{tab:resources}) provided quantitative estimates of the telescope time required to support cosmology measurements with LSST supernovae and strong-lens systems \cite{kavli}.

\begin{table}[htbp!]
    \centering
    \begin{tabular}{|c|c|}

        \hline
         Facility & Supernova single-object follow-up requirements  \\
         \hline
        4 m spectroscopy & 60--180 nights total \\
        8 m spectroscopy & 180--540 nights total \\
        $>20$m spectroscopy & 180--540 nights total \\ 
        \hline
        Facility & Strong Lensing single-object follow-up requirements  \\
        \hline
         2--4~m non-AO imaging & $<8000$ hours total \\
        $>8$m AO imaging & $\sim30$ hours, split amongst 8~m+ and 30~m+ telescopes\\
        $>8$m spectroscopy & $\sim100$ hours, split amongst 8~m+ and 30~m+ telescopes \\
         \hline
    \end{tabular}
    \caption{Summaries of the required resources, as estimated in the Kavli/NOAO/LSST study on Ground-based Optical/IR follow-up. \label{tab:resources}}
 
\end{table}

Although in many cases suitable instruments exist (e.g., Keck/OSIRIS) or are being developed (e.g., Gemini/SCORPIO), cases remain where the telescopes, both ground- and space-based, best positioned for this work have suboptimal instrumentation; as a result, additional development, not just telescope time, may be needed.

\newpage
\begin{center}
{\large\bf Acknowledgements}    
    
\end{center}
The LSST Dark Energy Science Collaboration acknowledges ongoing support from the Institut National de Physique Nucl\'eaire et de Physique des Particules in France; the Science \& Technology Facilities Council in the United Kingdom; and the Department of Energy, the National Science Foundation, and the LSST Corporation in the United States.  DESC uses resources of the IN2P3 Computing Center (CC-IN2P3--Lyon/Villeurbanne - France) funded by the Centre National de la Recherche Scientifique; the National Energy Research Scientific Computing Center, a DOE Office of Science User Facility supported by the Office of Science of the U.S.\ Department of Energy under Contract No.\ DE-AC02-05CH11231; STFC DiRAC HPC Facilities, funded by UK BIS National E-infrastructure capital grants; and the UK particle physics grid, supported by the GridPP Collaboration.  This work was performed in part under DOE Contract DE-AC02-76SF00515.

\newpage

\bibliographystyle{unsrt_truncate}
\bibliography{main}

\end{document}